\documentclass[doublecol]{epl2} 
\usepackage[english]{babel}
\usepackage{amsmath,
amssymb,amsfonts,latexsym}
\usepackage{graphicx}
\usepackage{color}
\usepackage{subfigure}
\usepackage{afterpage}
\usepackage{booktabs}
\usepackage{blkarray}
\usepackage{cite}
\usepackage{ulem}
\usepackage{bm}
\usepackage{braket}
\usepackage{nicefrac}
\usepackage{etoolbox}
\usepackage{mathrsfs} 
\usepackage{mathtools}
\usepackage[protrusion=true,expansion=true]{microtype}
\usepackage{amsmath}
\usepackage[usenames,dvipsnames,table]{xcolor}
\usepackage[colorlinks,
citecolor=blue,linkcolor=red,urlcolor=blue]{hyperref}
\usepackage{amsfonts,amsmath,amssymb}
\usepackage{soul} 
\usepackage{enumitem} 
\usepackage{blindtext} 
\definecolor{Blue}{rgb}{0.00, 0.00, 1.00}
\definecolor{Red}{rgb}{1.00, 0.00, 0.00}
\definecolor{labelkey}{cmyk}{.1,.7,0.5,0}

\newcommand{\de}{\partial}
\newcommand{\be}{\begin{equation}}
\newcommand{\ee}{\end{equation}}
\newcommand{\nn}{\nonumber\\[4pt]}
\newcommand{\qqq}{\end{document}}

\title{Enhanced correlations due to ballistic transport}
\shorttitle{Enhanced correlations due to ballistic transport} 
\author{D.~De Angelis$^1$   \and   J.~De Nardis$^1$ \and S.~Scopa$^{1,2}$ }
\shortauthor{D. De Angelis {\it et al.}}
\institute{                    
  \inst{1}{\small Laboratoire de Physique Th\'eorique et Mod\'elisation, CNRS UMR 8089,
		CY Cergy Paris Universit\'e, 95302 Cergy-Pontoise Cedex, France}\\	\inst{2}{\small Laboratoire de Physique de l’\'Ecole Normale Superieure, CNRS, ENS \& Universit\'e PSL,
Sorbonne Universit\'e, Universit\'e Paris Cit\'e, 75005 Paris, France}}
\vspace{-1cm}
\abstract{We investigate the nature of density-density correlations in a 1D gas of hard-core particles initially prepared at equilibrium (either at zero or finite temperature) on a semi-infinite line and subsequently let to expand into the other (initially empty) half of the system. Using a combination of analytical techniques based on exact methods and asymptotic hydrodynamic approaches, we discuss the behavior of the gas as its initial temperature varies, and back up our derivations with numerical exact diagonalization of the model. Our findings reveal that, irrespective of the initial temperature, the non-equilibrium behavior of density-density correlations at sufficiently large times is characterized by algebraic decay. Furthermore, we provide analytical results based on quantum generalized hydrodynamics that match with the numerical data both at zero and finite temperature.}

\begin{document}

\maketitle 
\section{Introduction}The decay type of 1D correlation functions unveils crucial insights about the system's nature. Algebraic decay signifies long-range correlations, often related to critical phenomena and conformal invariance \cite{difrancesco_book,henkel2013conformal}. In contrast, exponential decay indicates short-range correlations, typical of non-critical phases or disordered systems \cite{ziman1979models}. A paradigmatic example regards gapless 1D models, whose low-temperature properties can be described through Luttinger liquid theory and exhibit long-range correlations \cite{haldane1981luttinger, giamarchi2003quantum}. As the temperature rises, an exponential suppression of correlations in the gas is instead observed. In essence, at equilibrium, gapless 1D quantum models are expected to display either algebraic or exponential correlations depending on whether the state is at zero or finite temperature.\\
\indent
Out of equilibrium, this clear dichotomy disappears, and dynamics can give rise to long-range correlations in finite-temperature gases, e.g. \cite{Rigol2004quasicond,prosen2010longrange,vidmar_iyer_17,eisler2014area,takacs2024quasicondensation}. Among various quench dynamics, bi-partite protocols stand out as one of the simplest (yet nontrivial) setups for studying quantum transport \cite{antal1999transport,Collura_2018,Scopa2023domainwall}, and they will be the focus here. In these protocols, two gases are prepared at equilibrium with different chemical potentials and/or temperatures (and possibly other Lagrange multipliers in the integrable case) and then joined together to initiate non-equilibrium dynamics around the junction. Such protocols are not simply theoretical but they can be realised in cold atomic systems for example by shining lasers to remove part of the cloud, see \cite{Dubois2023}.  Recent theoretical advances in the theory of generalized hydrodynamics (GHD) have shown that the dynamic properties of local thermodynamic quantities in integrable gases, evolving unitarily from the bi-partite configuration, follow a ballistic scaling in $\sim x/t$ \cite{bertini2016transport,castro2016emergent}. However, it remains unclear what to expect for non-local objects such as two-point functions and entanglement, and how their behavior relates to the initial temperature. Some results for zero-temperature correlations and entanglement \cite{scopa2021exact,ruggiero2021quantum,Scopa_2023}, derived through the novel framework of quantum generalized hydrodynamics (QGHD) \cite{ruggiero2020quantum}, reveal that simple correlation functions, such as density-density and one-particle density matrix, display long-range correlations during quench dynamics, a behavior expected to hold true even for higher-order correlations. Zero-temperature entanglement shows a logarithmic growth in time \cite{ares2022entanglement,scopa2022exact,Scopa_2023}, typical of area-law phases, related to the expanding correlated region. At finite temperature, hydrodynamic equations for the correlation of conserved charges (e.g., density-density) can be derived through ballistic macroscopic fluctuation theory (BMFT) \cite{Doyon2023BMFT}, while volume-law entropy is predicted from Yang-Yang thermodynamics within the GHD framework \cite{Bertini2018,Capizzi_2023b}. Despite these partial results, modern literature lacks a comprehensive understanding of the interplay between temperature and the decay of system correlations in bi-partite configurations, which motivates our analysis below. \\
In this work, we focus on hard-core particles, and we derive analytical results supporting a $\sim 1/z^2$ decay at sufficiently large times and distance $z$, regardless of the value of the initial temperature. The latter typically characterize equilibrium zero-temperature states but not finite-temperature ones, where correlations are expected to decay exponentially (cf~paragraph below). Such surprising results was already pointed out in previous works, see e.g.~\cite{PanfilDeNardis2018,Doyon2015,Rigol1,Rigol2}, but here we give a full proof justification based on fluctuations on top of the hydrodynamic evolution.
\indent
\section{ Model and quench protocol}
We consider a 1D gas of non-interacting fermions with Hamiltonian
\be\label{eq:gas}
\hat{H}_{\mu,V}=\int  dx \ \hat\Psi^\dagger_x\left( -\frac{\de_x^2}{2}-\mu +  V(x)\right)\hat\Psi_x
\ee
where $\hat\Psi^\dagger_x$, $\hat\Psi_x$ are the creation, resp. annihilation, operators of a fermionic particle at position $x$, satisfying canonical on-site anticommutation relations $\{\hat\Psi^\dagger_x,\hat\Psi_y\}=\delta(x-y)$; $\mu$ is the chemical potential. For what follows, one can equivalently interpret the system as a gas of impenentrable bosons \cite{girardeau1960relationship}, and the Hamiltonian \eqref{eq:gas} as the result of a Jordan-Wigner transformation. We also consider the presence of a confining potential $V(x)$ coupling with the fermionic density. Indeed, interpreting the bi-partite initial configuration as an inhomogeneous equilibrium state will become useful in the following. Specifically, we design the trapping potential as a hard wall
\be\label{eq:hard-wall}
V(x)=\begin{cases} 0 \quad \quad \ \text{if $x\geq 0$},\\ +\infty \quad \text{otherwise},\end{cases}
\ee
so that the system is initially confined in the semi-infinite line $x\geq 0$ and otherwise empty. This choice leads to the state of the system at time $t=0$:
\be\label{eq:bipartite-init-state}
\hat\varrho(0)= \prod_{x<0}\ket{0}_x\bra{0}_x \ \otimes \ \hat\varrho^\text{GGE}_{\mu,T}
\ee
where $\ket{0}_x$ is the single-particle vacuum state, $\hat\Psi_x\ket{0}_x=0$, while 
\be
\hat\varrho^\text{GGE}_{\mu,T}= \frac{e^{-\hat{H}_{\mu,0}/T}}{{\cal Z}_{\mu,0}}, \quad {\cal Z}_{\mu,0}=\text{tr}\left(e^{-\hat{H}_{\mu,0}/T}\right),
\ee
is the Generalized Gibbs ensemble (GGE) \cite{rigol2008gge} for the Fermi gas at temperature $T$ and chemical potential $\mu$ occupying the right part of the system. Equivalently, one can write $\hat\varrho(0)=\exp(-\hat{H}_{\mu,V})/{\cal Z}_{\mu,V}$ such that the potential \eqref{eq:hard-wall} projects the l.h.s. onto the many-body vacuum.\\

At time $t>0$, the hard wall is switched off ($V=0$), and the system evolves unitarily as
\be\label{eq:rho-t}
\hat\varrho(t)= e^{-it \hat{H}_{\mu,0}}\ \hat\varrho(0)\ e^{it \hat{H}_{\mu,0}}.
\ee

During this quench dynamics, we shall focus on the evolution of the particle density
\be
\rho(x;t)=\braket{\hat\Psi^\dagger_x\hat\Psi_x}\equiv\text{tr}\bigl(\hat\varrho(t) \hat\Psi^\dagger_x\hat\Psi_x\bigr),
\ee
and of the connected density-density correlation
\be
{\cal C}(x,x';t)=\text{tr}\bigl(\hat\varrho(t) \hat\Psi^\dagger_x\hat\Psi_x \hat\Psi^\dagger_{x'}\hat\Psi_{x'}\bigr) - \rho(x;t)\,\rho(x';t).
\ee

\subsection{Initial correlations}
Given the non-interacting nature of the gas and the choice of potential \eqref{eq:hard-wall}, establishing the exact initial correlation functions in the gas becomes an easily attainable task by virtue of Wick's theorem. Defining the fermionic propagator of an infinite gas (i.e., translationally invariant~--~whose quantities are denoted hereafter with subscript $0$)  as
\be
G_0(x-x')=\braket{\hat\Psi^\dagger_x\hat\Psi_{x'}}=\int_{-\infty}^\infty \frac{dq}{2\pi}\, e^{iq(x-x')} \, n(q),
\ee
where
\be{
n(k)=\frac{1}{1+\exp\left[(k^2/2-\mu)/T\right]}}
\ee
is the Fermi-Dirac distribution, one has readily access to the density on the diagonal of the propagator, namely $\rho_0=G_0(0)$, and to the (connected) density-density correlation as
\be\label{eq:init-corr}
{\cal C}(x,x';0)=\delta(x-x')\, G_0(x-x') - |G_0(x-x')|^2.
\ee

In the ground state, the fermionic occupation $n(k)$ takes the form of a step function between the two Fermi points $\pm k_F=\pm\sqrt{2\mu}$, $n_\text{GS}(k)=\{ 1, \ \text{if} \ |k| \leq k_F;  \; 0\ \text{otherwise}\}$
and the previous result can be written explicitly, as 
\be\label{eq:corr-gs-in}
{\cal C}_\text{GS}(x,x';0)=\frac{-1+\cos\bigl(2k_F(x-x')\bigr)}{2\pi^2(x-x')^2}.
\ee

The algebraic decay $\sim1/|x-x'|^2$ {at large distances} in eq.~\eqref{eq:corr-gs-in} is a {renowned} result for the ground-state density-density correlations of 1D superfluids, obtainable through Luttinger liquid theory \cite{giamarchi2003quantum}.
 {Eq.~\eqref{eq:corr-gs-in} also accounts for short-wavelenght oscillatory terms, which will be neglected hereafter as they contribute only subleading corrections that do not affect the asymptotic power-law decay.} 
On the other hand, for a finite-temperature gas, eq.~\eqref{eq:init-corr} yields an exponential decay of correlations (for $x\neq x'$):
\be\label{eq:corr-th-in}
{\cal C}_\text{th}(x,x';0)=-|G_0(x-x')|^2\propto e^{-T|x-x'|/{\rho_0}}.
\ee
\begin{figure}[t]
\centering
\includegraphics[width=\columnwidth]{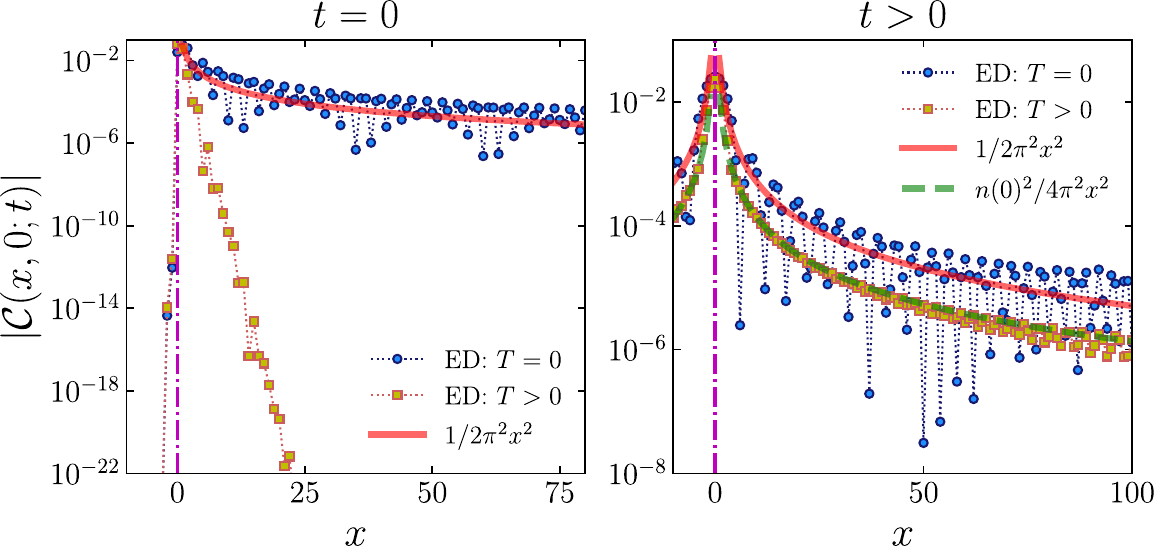}
\caption{Connected density-density correlation ${\cal C}(x,x';t)$ with $x'=0$ as function of $x$, for $t=0$ (left panel) and $t=100$ (right panel). Exact diagonalization numerical data (ED) is obtained for a free fermionic chain of size $L=2000$ with $\mu=0.5$, and temperature $T=0.5$ (squares) or zero (circles). Thick solid (resp.~dashed) lines show the analytical prediction $|{\cal C}(x,0;t)|\sim \frac{1}{2\pi^2x^2}$ (resp.~$|{\cal C}(x,0;t)|\sim \frac{n(0)^2}{4\pi^2 x^2}$) derived in the main text. Dot-dashed axes mark the junction position $x'=0$. }\label{fig:num-fact}
\end{figure}

\subsection{ Numerical evidence}As anticipated in the introduction, a qualitative change (from exponential to $\sim 1/|x|^2$) in the behavior of density-density correlations of the gas at finite temperature can be easily observed numerically, as shown in Fig.~\ref{fig:num-fact}. We shall complement this numerical evidence with analytical results. Although formal expressions can still be written down, obtaining explicit results becomes demanding in such non-equilibrium and non-homogeneous situations. Hence, we shall approach the calculation with hydrodynamic tools. \\

\section{Hydrodynamic approach}To begin with, we consider the Wigner operator of the free Fermi gas in absence of potentials $V=0$, defined as
\be
\hat{n}_0(x,k)=\int_{-\infty}^{\infty} d\xi \, e^{-ik\xi}\, \hat\Psi^\dagger_{x+\xi/2}\, \hat\Psi_{x-\xi/2}
\ee
and satisfying, by construction, $n(k)=\braket{\hat{n}_0(x,k)}$. The particle density is then simply obtained as
\be
\rho_0=\int_{-\infty}^\infty \frac{dk}{2\pi} \braket{\hat{n}_0(x,k)}.
\ee

We also introduce the (equal-time) dynamic structure factor as
\be
S_0(x,k;x',k')=\braket{\hat{n}_0(x,k)\,\hat{n}_0(x',k')},
\ee
or explicitly,
\begin{align}\label{eq:struc-fact}
S_0=\,&\delta(k-k')\int_{-\infty}^\infty \frac{dq}{2\pi}\int_{-\infty}^\infty \frac{dq'}{2\pi} \, e^{iq(x-x')}\,  e^{-iq'(x-x')}\nn
&\times n(q)\,[1-n(q')]\,\delta(k-\frac{q+q'}{2}),
\end{align}
such that the density-density correlation is given as
\be\label{eq:init-corr-S}
{\cal C}(x,x')=\int_{-\infty}^\infty \frac{dk}{2\pi} \int_{-\infty}^\infty \frac{dk'}{2\pi} S_0(x,k;x',k').
\ee
It is easy to show that eq.~\eqref{eq:init-corr-S} is equivalent to \eqref{eq:init-corr}. By recasting $S_0$ as ($z=x-x'$),
\begin{align}
&S_0(x,k;x',k')=\delta(k-k')\,e^{2ikz}\,G_0(2z)\nn
& -\delta(k-k')\int d\xi\, e^{-2i k\xi}\, G_0(z+\xi/2)\,G_0^*(z-\xi/2),
\end{align}
one can see the second term is projected to $\xi=0$ by the integration over momenta, and thus eq.~\eqref{eq:init-corr} is recovered.\\

Extending these expressions to the bi-partite state \eqref{eq:bipartite-init-state} is quite simple. We write the Wigner function as 
\be\label{eq:n-in-cond}
{n}(x,k; 0)= \Theta(x)\, {n}(k),
\ee
and the structure factor
\be
S(x,k; x',k'; 0)= \Theta(x)\, \Theta(x')\, S_0(x,x';k,k'),
\ee
where $\Theta(\cdot)$ is the Heaviside step function, physically enforcing a zero density (resp. correlation) on the left part of the system at $t=0$.\\

\subsection{Hydrodynamic evolution} 
\begin{figure}[t]
\centering
\includegraphics[width=.9\columnwidth]{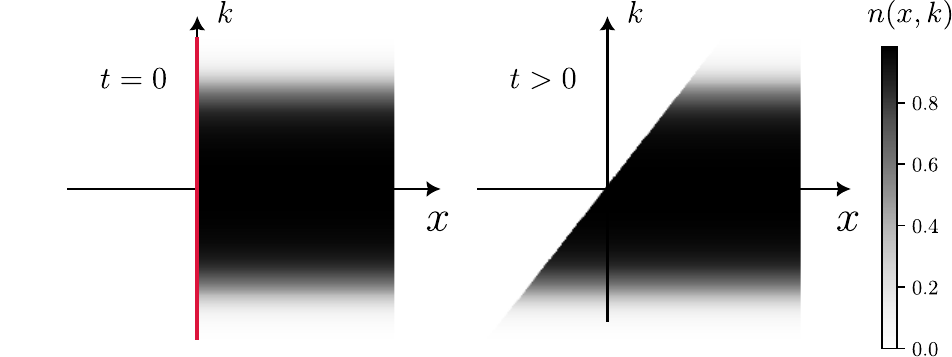}
\caption{Illustration of the Wigner function $n(x,k;t)$ at $t=0$ (left) and for $t>0$ (right), obtained from eq.~\eqref{eq:euler-n} with initial condition \eqref{eq:n-in-cond}. In the illustration, we set $\mu=0.5$ and $T=0.5$.}\label{fig:sketch}
\end{figure}

\begin{figure}[t]
\centering
\includegraphics[width=.9\columnwidth]{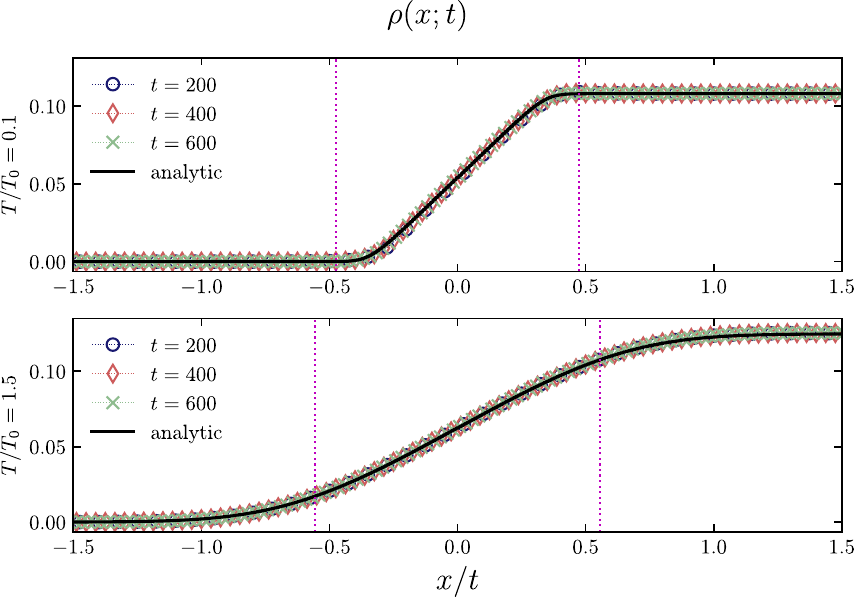}
\caption{Particle density as function of $x/t$ for a system with $\mu=0.5$ and different values of temperature $T/T_0$ (rows). Here $T_0=k_F^2$ gives an estimate of the degenerate temperature of the gas \cite{Bouchoule2011}. Symbols are numerical data at different times obtained with exact diagonalization, while solid lines show the prediction of \eqref{eq:dens-t}. Vertical axes mark the location of $\pm k_*$ in \eqref{eq:lightcone}.}\label{fig:density}
\end{figure}

At $t>0$, the time-evolution of the Wigner function is given by the Moyal equation \cite{moyal1949quantum,fagotti2020locally,fagotti2017higher}, and, at lowest order in the $\de_x$ and $\de_k$ derivatives, it satisfies the Euler equation
\be\label{eq:euler-n}
(\de_t+ k \de_x)\,n(x,k;t)=0
\ee
solved by $n(x,k;t)=n(x-kt,k;0)$. An illustration of the dynamics of $n(x,k;t)$ for the bi-partitioning protocol is given in Fig.~\ref{fig:sketch}; eq.~\eqref{eq:euler-n} yields the result for the particle density
\be\label{eq:dens-t}
\rho(x;t)=\int_{-\infty}^\infty \frac{dk}{2\pi}\, \Theta(x-kt)\, n(k),
\ee
characterized by the scaling $x/t$ as shown in Fig.~\ref{fig:density}. Notice that the dynamics of the density is characterized by a propagating front, often referred to as \textit{light cone}. At zero temperature, one has $x/t=\pm k_F$, while, for the thermal gas, one can define a light-cone region as $x/t=\pm k_*$, where $k_*$ is estimated (for not too high temperatures) as the half-width at half-height of the Fermi-Dirac distribution, namely,
\be\label{eq:lightcone}
k_*\simeq \sqrt{2\mu-2T\log\left(\frac{e^{\mu/T}}{2+e^{\mu/T}}\right)}.
\ee


From hydrodynamic projection in BMFT, it has been recently shown that a similar hydrodynamic equation is satisfied by the equal-time structure factor \cite{Doyon2023BMFT}:
\be
(\de_t + k \de_x  + k' \de_{x'})\, S(x,k;x',k';t)=0,
\ee
solved by $S(x,k,x',k';t)=S(x-kt,k; x'-k't,k';0)$. This gives for the density-density correlation
\begin{align}\label{eq:corr-therm-ex}
{\cal C}(x,x';t)=&\iint \frac{dk\, dk'}{4\pi^2}\Theta(x-kt)\Theta(x'-k't)\nonumber\\ &\quad\times S_0(x,k,x',k')
\end{align}
Notice that, due to rapidly oscillating exponentials and multiple integrations, the numerical evaluation of this formal result is challenging.
\subsection{Emergence of algebraic decay} 
\begin{figure}[h]
\centering
\includegraphics[width=.85\columnwidth]{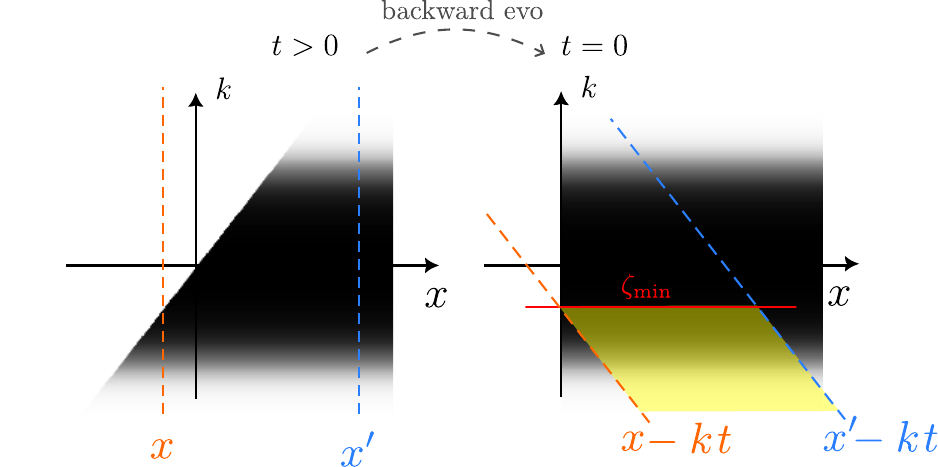}
\caption{Correlated domain contributing to ${\cal C}(x,x';t)$: the time evolution generates an effective Fermi point at $\zeta_\text{min}=\min(x,x')/t$, from whose vicinity the dominant contribution to the density-density correlation is coming, {namely when $k \approx \zeta_{\text{min}}$}.} \label{fig:sketch2}\vspace{.2cm}
\end{figure}

By inspecting the structure of eq.~\eqref{eq:corr-therm-ex}, we notice that the single-particle propagators with momenta $q$, $q'$ entering \eqref{eq:struc-fact} must satisfy
\be
k\equiv k'= \frac{q+q'}{2}\leq \min(x,x')/t\equiv \zeta_\text{min}.
\ee
This condition is illustrated in Fig.~\ref{fig:sketch2}, and it has a clear physical interpretation. Density-density correlations are given by the integral sum over $k$ of the product of two propagators carrying momenta $q$ and $q'$, such that their average equals $k$ (cf. eq.~\eqref{eq:struc-fact}). Through ballistic transport, particles initially occupying the bulk region of $n(x,k)$ are {outrun} by the thermal tails, and form this way an effective Fermi point $\zeta_\text{min}$ for the gas. It becomes evident that the integrals in \eqref{eq:corr-therm-ex} are dominated by the contribution where $q\approx q'\simeq \min(x,x')/t$, and are otherwise exponentially suppressed. Therefore, for sufficiently large times, we can approximate the density-density correlation \eqref{eq:corr-therm-ex} as
{
\begin{align}\label{eq:corr-therm-app}
{\cal C}(x,x';t)\approx\ &\frac{n(\zeta_\text{min})^2}{4\pi^2}\left[\int dq\ e^{iq(x-x')} \right]_{q= \zeta_\text{min}}\nn
&\times \left[\int dq'\ e^{-iq'(x-x')}\right]_{q'=\zeta_\text{min}}\approx  \frac{n(0)^2}{4\pi^2(x-x')^2}.
\end{align}
}
This estimate is compared against numerical data in Fig.~\ref{fig:num-fact}. Despite its simplicity, it provides an accurate description of the density-density correlation out of equilibrium. And most importantly, it confirms the development of long-range order $\sim 1/|x-x'|^2$, offering a physical explanation for its onset (cf. Fig.~\ref{fig:sketch2}). A similar development of long-range correlations is also found in domain-wall-like initial states see e.g.~\cite{takacs2024quasicondensation,Rigol1,Rigol2}, where spatial correlations are initially zero and develop due to dynamics.  Notably, performing the same approximation at zero temperature would yield a prefactor of $1/(2\pi^2)$ in \eqref{eq:corr-therm-app}, as two Fermi points would now contribute to the density-density correlation.

\section{Exact result at zero temperature}At zero temperature, exact results for the density-density correlation can be determined using quantum fluctuating hydrodynamics \cite{ruggiero2020quantum,scopa2021exact,ruggiero2021quantum,ares2022entanglement,scopa2022exact,Scopa2023domainwall}. Let us begin with the case of an infinite system at equilibrium, for which a standard procedure applies \cite{haldane1981luttinger,giamarchi2003quantum}. The first step consists in treating the particle density $\rho(x)$ as an operator, by re-introducing the leading quantum fluctuations in terms of density fluctuating fields $\delta\hat{\rho}_\sigma(x)$ around the two Fermi points $\sigma=\pm$. Specifically,
\be\label{eq:dens-exp}
\hat{\rho}(x)\simeq  \rho(x) +\sum_{\sigma=\pm} \delta\hat\rho_\sigma(x)
\ee
valid at leading order in the Haldane's expansion, see e.g. \cite{giamarchi2003quantum}. Using that
\be
\braket{\delta\hat\rho_\sigma(x)\,\delta\hat\rho_{\sigma'}(x')}=\frac{-1}{4\pi^2} \frac{\delta_{\sigma\sigma'}}{{\rm d}(x-x')^2},
\ee
we readily recover the formula for the connected density-density correlation \eqref{eq:corr-gs-in},
\be\label{eq:corr-fluct-inf}
{\cal C}(x,x';0)\simeq \sum_{\sigma} \braket{\delta\hat\rho_\sigma(x)\,\delta\hat\rho_{\sigma}(x')} =\frac{-1}{2\pi^2\,{\rm d}(x-x')^2},
\ee
up to density ripples generated from higher-order corrections in the density operator expansion \eqref{eq:dens-exp}, which we hereafter neglect (see e.g., Refs.~\cite{brun2018inhomogeneous,kheruntsyan-shock,Urilyon2024} for a discussion). Here ${\rm d}(\cdot)$ is a distance function: ${\rm d}(x)=x$ for the infinite system, while it becomes equal to the chord distance ${\rm d}(x)=2R\sin(\frac{\pi x}{2R})/\pi$ in a system of finite size $R$.\\
\begin{figure}[h]
\centering
\includegraphics[width=.7\columnwidth]{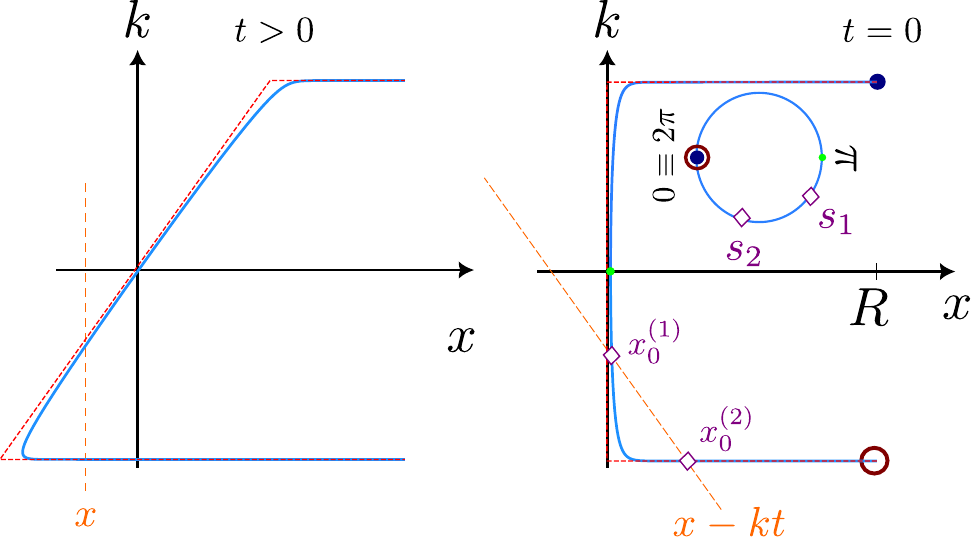}
\caption{Schematic procedure for finding the coordinates $s_{1,2}$ entering in \eqref{eq:qghd-dd}: for a given position $x$ and time $t>0$, one can determine the intersections $x_0^{(1,2)}$ on the initial Fermi contour by solving \eqref{eq:roots}. These roots correspond to $s_{1,2}$ on the unit circle as defined by \eqref{eq:s-coord-def}. The solid blue curve denotes the regularized Fermi surface with potential \eqref{eq:exp-reg} while the red dashed line shows the corresponding hard-wall limit attained for $\alpha\to\infty$.
}\label{fig:reg_sketch}
\end{figure}
We now move on to the case of interest, whose calculation is made possible using the QGHD framework \cite{ruggiero2020quantum,scopa2021exact}. As noted already in Ref.~\cite{scopa2021exact}, the case of the hard wall \eqref{eq:hard-wall} on the semi-infinite line is pathological, and thus requires a regularization to be treated with hydrodynamic approaches. We shall consider a regularization of the box potential \eqref{eq:hard-wall} as
\be\label{eq:exp-reg}
V(x)=e^{-\alpha x}, \qquad -R\leq x \leq R,
\ee
and we further take the limit $\alpha, R \to\infty$ to recover the case of our interest.  A general formula valid for release from arbitrary (smooth enough) potentials $V(x)$ has been derived in Ref.~\cite{ruggiero2021quantum}, and reads as
\be\label{eq:qghd-dd}
{\cal C}_\text{GS}(x,x';t)=\frac{-1}{4\pi^2}\sum_{a,b=1}^2 \frac{ds_a}{dx} \frac{ds'_b}{dx'}  \frac{1}{{\rm d}(s_a-s'_b)^2}.
\ee
This formula appears as a simple extension of eq.~\eqref{eq:corr-fluct-inf}, which encodes all details on the quench protocol in the choice of the coordinates $s_{1,2}\equiv s(x_0^{(1,2)})$. The latter are determined with the following procedure:\\[3pt]
1.~For given $x$ and $t$, find the two solutions $x_0^{(1,2)}$ of
\be\label{eq:roots}
\frac{(x-x_0)^2}{2t^2}=\mu- V(x_0).
\ee
2.~Given a root $x_0\equiv x_0(x,t)$ of eq.~\eqref{eq:roots}, define the angular coordinate \cite{scopa2021exact}
\begin{align}
 \vartheta(x_0)=&\frac{\pi}{\cal N}\int_{x_0}^{R} \frac{dx}{\sqrt{2(\mu-V(x))}}
\nn
& =\frac{\pi}{\cal N} \sqrt{\frac{2}{\mu\alpha^2}}\tanh^{-1}\sqrt{1-\frac{e^{-\alpha x}}{\mu}}\Bigg\vert_{x_0}^{R},
\end{align}
such that $0\leq\vartheta(x_0)<\pi$, with normalization 
\be
{\cal N}=\sqrt{\frac{2}{\mu\alpha^2}}\tanh^{-1}\sqrt{1-\frac{1}{\mu} e^{-\alpha R}}.
\ee
3.~From this, one can obtain the coordinate
\be\label{eq:s-coord-def}
s(x_0)=\begin{cases}\vartheta(x_0) \quad \text{if $\text{sign}(x-x_0)\geq 0$,}\\[6pt]
2\pi-\vartheta(x_0) \quad \text{otherwise.}
\end{cases}
\ee
\begin{figure*}[t]
\centering
\includegraphics[width=\textwidth]{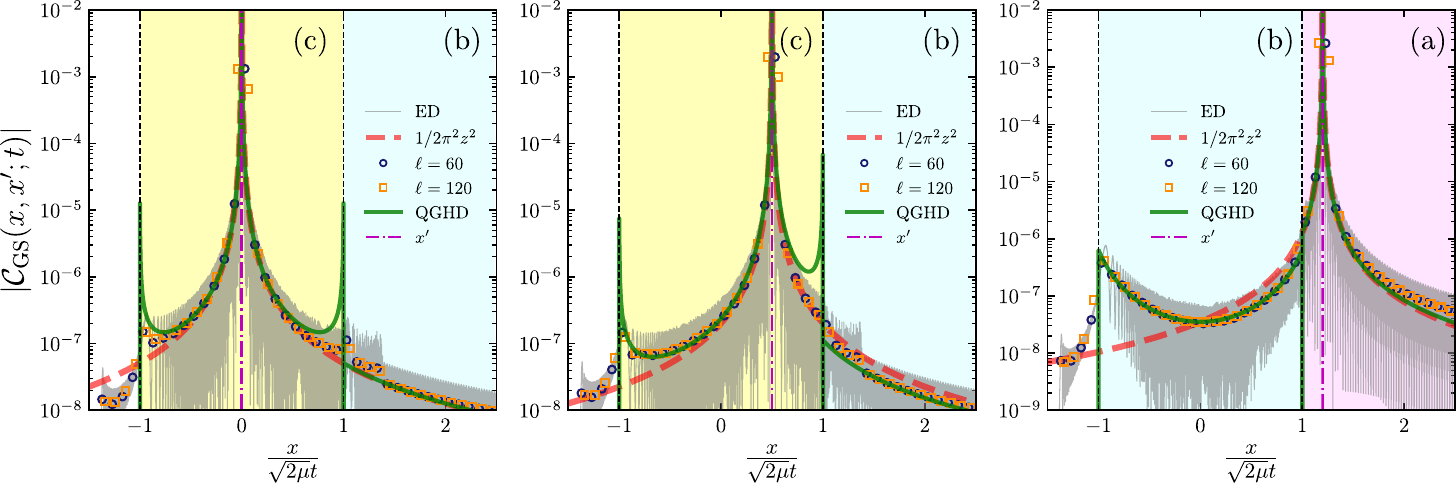}
\caption{Connected density-density correlation as a function of $x$ for different values of $x'$ (shown in different panels), at $t/R=0.2$ after the release. The system is initially in the ground state of \eqref{eq:hard-wall}. Thin solid lines represent numerical data obtained with exact diagonalization (ED), while symbols indicate the spatial average of the data over an interval of length $\ell$. The solid thick curve is the QGHD prediction from \eqref{eq:corr-qghd}, while the thick dashed line represents the estimate $1/2\pi^2z^2$ ($z=x-x'$). We set $\mu=0.5$ (and $R=5000$ for the numerical data). In each panel, the regions (a)-(c) of Eq.~\eqref{eq:corr-qghd} are highlighted in different colors.}\label{fig:qghd-res}
\end{figure*}
\begin{figure}[h]
\centering
\includegraphics[width=\columnwidth]{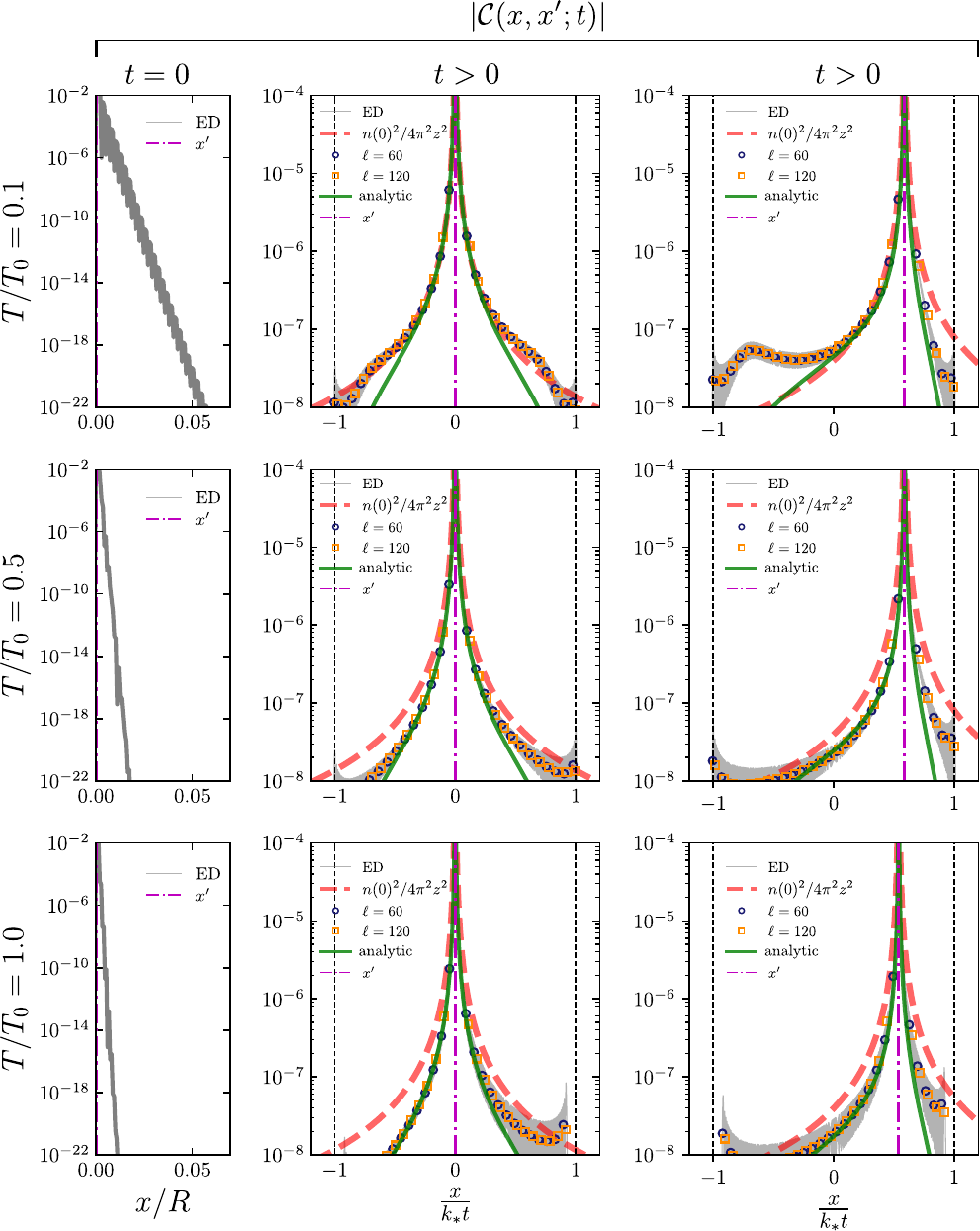}
\caption{Connected density-density correlation as a function of $x$ for different values of $x'$ (different panels), at $t=0$ (leftmost panel) and $t/R=0.4$ after the release (right panels). The system is initially at equilibrium with temperature $T$. Thin solid lines represent the exact diagonalization numerical data (ED), with symbols indicating the spatial average of ED data over an interval of size $\ell$. The solid thick curve is the conjecture \eqref{eq:corr-qghd-T}, and the dashed thick line represents the estimate $n(0)^2/4\pi^2 z^2$ ($z=x-x'$) in Eq.~\eqref{eq:corr-therm-app}. We set $\mu=0.5$, $T_0=2\mu$ (and $R=5000$ for the numerical data).}\label{fig:qghd-res-T}
\end{figure}
This procedure, illustrated in Fig.~\ref{fig:reg_sketch}, physically consists in mapping the Fermi points of the system at position $x$ and time $t$ onto a unit circle, with distances measured according to the propagation time of low-energy excitations on the initial inhomogeneous Fermi surface, see Refs.~\cite{brun2017one,brun2018inhomogeneous,scopa2020one} for details. In the hard-wall limit $\alpha\to\infty$, approximate analytical solutions to eq.~\eqref{eq:roots} can be found. In particular,\\[4pt]
\indent
-~if $x/t\geq \sqrt{2\mu}$, i.e., $x$ is outside the light cone but on the filled region, then one has the two roots $x_0^{(1,2)}= x\mp \sqrt{2\mu} t$, up to exponentially small corrections;\\[8pt]
\indent 
-~if $|x|/t\leq \sqrt{2\mu}$, i.e., $x$ is inside the light cone, then one still finds the root $x_0^{(2)}=x-\sqrt{2\mu} t$, together with the non-trivial root $ x_0^{(1)}=\log(\mu -\frac{x^2}{2t^2})/\alpha$ in a small region $\sim 1/\alpha$ near the wall, and up to algebraic corrections;\\[8pt]
\indent
-~if $x/t<-\sqrt{2\mu}$ then there are no roots.\\[4pt]
Moreover, the normalization becomes ${\cal N}\approx R/\sqrt{2\mu}$ for $\alpha\to\infty$, because the region where $V$ is non-zero shrinks near $x=0$ and asymptotically approaches zero measure. Depending on $x$ and $t$, we thus find two of the following three values for the $s$ coordinate
\be
s(x_0)=\begin{cases}
\pi(1-\frac{x+\sqrt{2\mu}t}{R}) \qquad \text{if $x/t\geq -\sqrt{2\mu}$}, \\[4pt]
\pi(1+\frac{x-\sqrt{2\mu}t}{R}) \qquad \text{if $x/t> \sqrt{2\mu}$},\\[4pt]
\pi-\frac{2\pi}{\alpha R} \tanh^{-1}(\frac{x}{\sqrt{2\mu}t}) \qquad \text{if $|x|/t\leq \sqrt{2\mu}$}.
\end{cases}
\ee

Inserting these values for $s$ in eq.~\eqref{eq:qghd-dd}, after some simple algebra, one finds for $\alpha, R\to\infty$
\be\label{eq:corr-qghd}
{\cal C}_\text{GS}(x,x';t)=\begin{dcases}
\text{(a):}\quad \frac{-1}{2\pi^2(x-x')^2}+\frac{-1}{2\pi^2(x+x')^2},\\[4pt]
\quad \text{if $x,x'> \sqrt{2\mu}t$;}\\[8pt]
\text{(b):}\quad \frac{-1}{4\pi^2(x-x')^2}+\frac{-1}{4\pi^2(x+x)^2},\\[4pt]
\quad \text{if $|x|\leq \sqrt{2\mu}t$ and $x'>\sqrt{2\mu}t$ }\\
\qquad \text{or}\\
\quad  \text{ $|x'|\leq \sqrt{2\mu}t$ and $x>\sqrt{2\mu}t$;}\\[8pt]
\text{(c):}\quad \frac{-1}{4\pi^2(x-x')^2}+\\
\frac{-1}{8\pi^2\mu t^2}\left[\left(1-\frac{x^2}{2\mu t^2}\right)\left(1-\frac{x'^2}{2\mu t^2}\right)\right]^{-1}\\
\times\left[\tanh^{-1}(\frac{x}{\sqrt{2\mu}t})-\tanh^{-1}(\frac{x'}{\sqrt{2\mu}t})\right]^{-2}\\[4pt]
\quad \text{if $|x|,|x'|\leq \sqrt{2\mu}t$,}
\end{dcases}
\ee
and otherwise zero. In Fig.~\ref{fig:qghd-res}, we test this formula against numerical data and find an excellent agreement in all regions (a)-(c), except very close to the light cone. In these regions, an ultraviolet regularization is needed to correct the behavior of the jacobians $ds(x_0)/dx$, see e.g. Ref.~\cite{scopa2021exact,Scopa_2023}.

\section{Improved result at finite temperature}Finally, by exploiting the exact solution \eqref{eq:corr-qghd} at zero temperature, and having understood the physical mechanism behind the formation of long-range order in thermal gases, we aim to refine the approximation in eq.~\eqref{eq:corr-therm-app}.
By isolating the dynamical contribution in \eqref{eq:corr-qghd} from those already present in the initial state, and by replacing the lightcone $k_F=\sqrt{2\mu}\to k_*(\mu,T)$, one finds that the density-density correlation for $|x|/t,|x'|/t\leq k_*$ is given by
\begin{align}\label{eq:corr-qghd-T}
{\cal C}_\text{th}(x,x';t)=&\frac{n(0)^2}{4\pi^2 k_*^2 t^2}\left[\left(1-\frac{x^2}{k_*^2t^2}\right)\left(1-\frac{x'^2}{k_*^2t^2}\right)\right]^{-1} \nn
&\times \left[\tanh^{-1}\left(\frac{x}{k_* t}\right)-\tanh^{-1}\left(\frac{x'}{k_* t}\right)\right]^{-2},
\end{align}
and exponentially suppressed otherwise. As shown in Fig.~\ref{fig:qghd-res-T}, this conjecture provides better agreement with the numerics than eq.~\eqref{eq:corr-therm-app}.

{Interestingly, the agreement of eq.~\eqref{eq:corr-qghd-T} with the data improves with temperature, but still for values of $T$ not too large. Indeed, we expect the zero-temperature result to give a more accurate prediction rather than eq.~\eqref{eq:corr-qghd-T} at $T\ll T_0 \sim k_F^2$, possibly with low-temperature corrections. Similarly, we expect that at large temperatures, thermal fluctuations extend to the bulk of the system, thus invalidating our mechanism, based on the dominant contribution coming from the effective Fermi point. To sum up, in case of small temperatures (cf. 1st panel of Fig.~\ref{fig:qghd-res-T}), eq.~\eqref{eq:corr-qghd-T} is suboptimal but it is possible to incorporate the small temperature effect into Luttinger liquid theory, as discussed in recent work~\cite{new-paper-rigol}.}\\

\section{Outlooks and conclusions}~
We investigated the asymptotic behavior of nonequilibrium density-density correlations of a 1D gas of hard-core particles during the free expansion from a bi-partite configuration with r.h.s. occupied by an equilibrium state (at either zero or finite $T$) and the l.h.s. is initially empty (eq.~\eqref{eq:bipartite-init-state}). For this quench protocol, we derived analytical results supporting the algebraic decay $\sim 1/|x-x'|^2$, observed numerically at sufficiently large distance $|x-x'|$ and time. Our analysis clarifies the physical mechanism through which an effective Fermi point is generated by the ballistic transport of the finite-temperature gas, and thus the onset of long-range correlations during the dynamics. Moreover, we exploited this effective Fermi surface to extend the QGHD result (eq.~\eqref{eq:corr-qghd}) to the finite temperature case. The resulting ansatz, eq.~\eqref{eq:corr-qghd-T}, is able to match the numerical data for a wide range of temperatures, see Fig.~\ref{fig:qghd-res-T}. We expect to observe the same enhancement of correlations in the Bose gas at finite interaction since a similar hydrodynamic argument applies. We plan to investigate the finite interaction case and the connection with experiments in future publications.\\

{\it Acknowledgments}. This work has been partially funded by the ERC Starting Grant 101042293 (HEPIQ)
(all authors); ANR-22-CPJ1-0021-01 (JDN), and MSCA Grant 101103348 (GENESYS) (SS). The authors acknowledge A Takacs, A Urilyon, J Dubail, P Ruggiero and I Bouchoule, for useful discussions and closely related collaborations.\\[3pt]
{\small [{This work has been partially funded} by the European Union. Views and opinions expressed are however those of the author(s) only and do not necessarily reflect those of the
European Union or the European Commission. Neither the European Union nor the European Commission can
be held responsible for them.]}

\bibliographystyle{eplbib}
\end{document}